\documentclass[12pt,english,a4paper]{iopart}
\usepackage[T1]{fontenc}
\usepackage[latin1]{inputenc}
\usepackage{iopams}
\usepackage{babel}
\usepackage{graphicx}

\begin{document}

\title{
  Microscopic modeling of a spin crossover transition}

\author{Harald O. Jeschke$^1$, L. Andrea Salguero$^1$, Badiur Rahaman$^2$, Christian Buchsbaum$^3$, Volodymyr Pashchenko$^3$, Martin U. Schmidt$^3$, Tanusri Saha-Dasgupta$^2$, Roser Valent{\'\i}$^1$}

\address{$^1$ Institut f{\"u}r Theoretische Physik, Universit{\"a}t Frankfurt,
Max-von-Laue-Str. 1, 60438 Frankfurt, Germany}

\address{$^2$ S.N. Bose National Centre for Basic Sciences,
JD Block, Sector 3, Salt Lake City, Kolkata 700098, India.}

\address{$^3$ Institut f\"ur Anorganische und Analytische Chemie, Universit{\"a}t Frankfurt,
Max-von-Laue-Str. 7, 60438 Frankfurt, Germany}

\ead{jeschke@itp.uni-frankfurt.de}

\begin{abstract}
In spin crossover materials, an abrupt phase transition between a
low spin state and a high spin state can be driven by temperature,
pressure or by light irradiation. Of a special relevance are Fe(II) based
coordination polymers where, in contrast to molecular systems, the
phase transition between a spin $S=0$ and a spin $S=2$ state
shows a pronounced hysteresis which is desirable for
technical applications. A satisfactory microscopic explanation of this
large cooperative phenomenon has been sought for a long time. The lack
of X-ray data has been one of the reasons for the absence of
microscopic studies. In this work, we present an efficient route to
prepare reliable model structures and within an {\it ab initio} density
functional theory analysis and effective model considerations we show
that in polymeric spin crossover compounds magnetic exchange between
high spin Fe(II) centers is as important as elastic couplings for
understanding
the phase transition. 
We discuss the relevance of these interactions for the cooperative behavior in these materials.   
\end{abstract}



\submitto{\NJP}

\maketitle

\section{Introduction}\label{sec:introduction}

An intensively debated class of materials with potential applications
as optical switches, sensors or memory
devices~\cite{Cobo06,Freysz05,Letard04,Guetlich02} are spin-crossover
polymer (SCP) systems involving transition metal ions linked with
organic ligands~\cite{Kahn98}. These systems show a sharp transition
triggered by variation of temperature, pressure or by light
irradiation between a low-spin (LS) ground state and a high-spin (HS)
excited state with a thermal hysteresis
loop~\cite{Guetlich94}. Especially important in these materials is the
large cooperativity shown at the HS-LS transition in contrast to
molecular spin crossover systems, for example.  The origin of this transition
and its cooperativity has been mainly discussed in the frame of
elastic
models~\cite{Guetlich94,Willenbacher88,Spiering89,Spiering04,Nishino07},
and only recently a possible role of magnetic exchange was
suggested~\cite{Timm05,Timm06}.  Still, a conclusive {\it ab initio}
microscopic study where all important interactions are considered is
missing and the origin of the large cooperativity has not been
completely settled. It is our purpose to investigate this issue in
what follows. {\it Ab initio} theoretical studies for SCP systems are
faced with major difficulties due to the nonexistence of
well-determined crystal structures.  To our knowledge, electronic
structure calculations have only been performed at the level of
semiempirical extended H\"uckel approximation for an idealized
triazole-bridged Fe(II) chain~\cite{Yoshizawa97}.  In the present
work, we overcome the unavailability of structural data by predicting
a crystal structure for a Fe(II) spin-crossover polymeric
crystal using known experimental constraints and a combination of
classical force field and quantum mechanical molecular dynamical
methods.  We analyze with Density Functional Theory (DFT) calculations
the LS-HS phase transition and show that there exists
an interplay between magnetic exchange and elastic properties that
is responsible for the large cooperativity in these systems.  We also
corroborate the quality of our designed structure by comparing with
magnetic experiments done on a real sample.
Our methodology and results  provide a new perspective on
 the parameters underlying  the traditional theoretical
approaches.

There have been a number of attempts to theoretically account for the
features of the HS-LS transition in spin crossover materials. 
Most of the theoretical work is based on elastic considerations. 
Two types of elastic models that focus on the faithful reproduction of
macroscopic quantities like the HS fraction, are the following:  In the
first approach, the cooperativity in the
HS-LS transition is defined in terms of local distortions which
interact with one another elastically causing a long range effective
interaction between spin states. This leads to an Ising-type
Hamiltonian~\cite{Wajnflasz71,Boukheddaden07a,Boukheddaden07b} $H = 
 \sum_{i,j} \tilde{J}_{ij} 
  \sigma_i \sigma_j $
which describes the elastic interaction between spin states (LS and
HS) in terms of fictitious spin operators ($\sigma$ = -1 (1) for LS
(HS)) coupled via a nearest neighbor interaction $\tilde{J}_{ij}$. The
coupling constants $\tilde{J}_{ij}$ are parameters of the theory and have not
yet been determined from a microscopic model. In an alternative
approach, the free energy of spin crossover systems is calculated
based on an anisotropic sphere model that describes volume and shape
changes of the lattice at the
transition~\cite{Willenbacher88,Spiering89,Spiering04}.
One of the few attempts to include magnetic interactions is the recent  proposal
done in Ref.~\cite{Timm06}  where the author considers
a one-dimensional model for HS-LS systems which contains elastic and magnetic Ising
exchange interactions. The ground-state phase diagram is then obtained
by the transfer matrix technique for different relative elastic to magnetic coupling
strengths.

In the  present work  we concentrate on the  electronic and magnetic
degrees of freedom in a spin crossover polymer, and investigate
 their influence on the microscopic origin of the cooperativity in the HS-LS
transition.  While it has been assumed in the past that the Fe(II)
center nearest neighbor interaction is mostly of phononic origin,
 our study indicates that a significant part of this
interaction arises from magnetic exchange.


\section{Crystal Structure}\label{sec:structure}

\begin{figure}
\begin{center}
\includegraphics[bb=72 366 393 700,clip,width=0.9\textwidth]{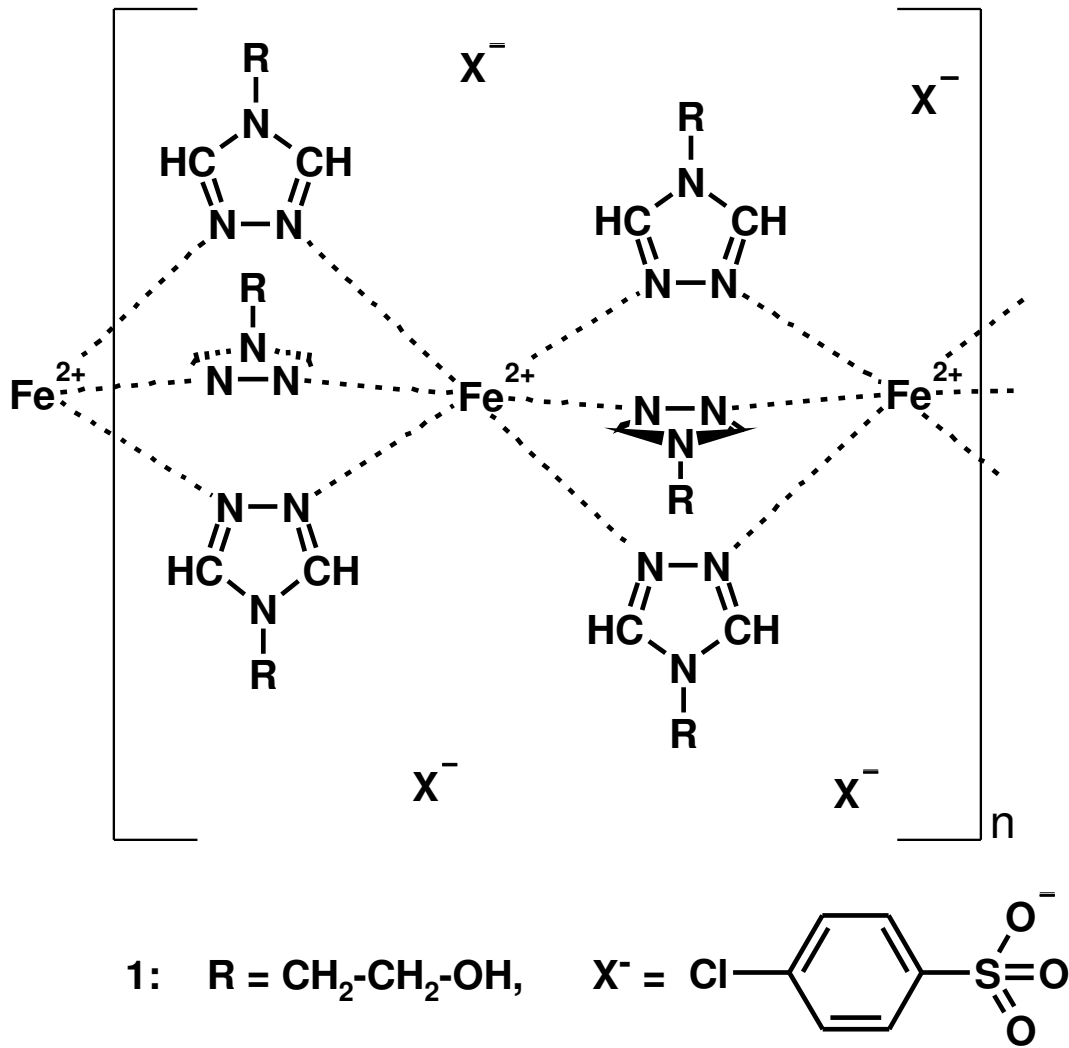}
\end{center}
\caption{
{\bf Chemical diagram of polymeric Fe(II)-triazole}. The
model compound {\bf 2} differs from the synthesized
compound {\bf 1} in the simplified R and X$^-$ groups.  }
\label{fig:triazole}
\end{figure}

For this investigation it is indispensable to obtain a reliable
crystal structure suitable for DFT analysis. We aim at describing the
complex Fe[(hyetrz)$_3$](4-chlorophenylsulfonate)$_2 \cdot
3$H$_2$O~\cite{Garcia04} ($hyetrz$ stands for
$4-(2'-hydroxyethyl)-1,2,4-triazole$) (see compound {\bf 1} in
Fig.~\ref{fig:triazole}), which was synthesized from
2-hydroxyethyltriazole and iron(II)-p-chlorobenzenesulfonate as
described in Ref.~\cite{Garcia97a}.

This compound precipitates as a fine, polymeric powder and single
crystals cannot be grown because the polymer is insoluble in water and
organic solvents~\cite{Guetlich04}. Melting or sublimation attempts
result in decomposition. Hence, X-ray structure analysis is not
possible and even the X-ray powder diagram consists of a few broad
peaks only which prevent the structure to be determined from X-ray
powder data. Other polymeric Fe-triazole compounds have similar
properties and no single crystal structure for these systems is
known. In the literature only single crystal structures for trimeric
Fe compounds (e.g. Ref.~\cite{Garcia00}), and polymeric
Cu-triazoles~\cite{Garcia03,Garcia97} were found. In these compounds,
the metal ion is coordinated by six nitrogen atoms and neighboring
metal ions are connected by three pyrazole bridges. For the polymeric
Fe triazoles a similar structure is assumed, see
Fig.~\ref{fig:triazole}. This structure is also supported by
spectroscopic methods~\cite{Garcia97} including solid state
NMR~\cite{Glaubitz07}.

In view of the above and the requirement of having reliable crystal
structures for microscopic studies, we design on the computer a model system of
polymeric Fe triazole, as a basis for calculating the electronic and
magnetic properties. We consider all experimental information
available and construct a crystal structure as close as possible to
the actual structure. We employ a method that has been developed and
tested on the coordination polymer
Cu(II)-2,5-bis(pyrazole-1-yl)-1,4-dihydroxybenzene which has a simpler
structure and less atoms per unit cell than polymeric
Fe-triazole~\cite{Jeschke07,Salguero07}.

Since we aim at understanding the HS-LS transition with accurate
all-electron DFT calculations, which are computer intensive, we keep
the essential features of the material and simplify those elements
that are secondary to the transition, like the nature of the
substituent R and the counter ion X$^-$ (spin transitions are observed
for a wide range of different substituents R and counter ions
X$^-$). We consider compound {\bf 2} in Fig.~\ref{fig:triazole} with R
= CH$_3$ and X$^-$ = F$^-$. This model structure has the short range
environment of the Fe(II) centers exactly as in structure {\bf 1} in
Fig.~\ref{fig:triazole} while the longer range environment ($\gtrsim
5$~{\AA}) of the Fe$^{2+}$ centers is significantly simplified. The
molecular geometry of compound \textbf{2} was constructed according to
data from single crystal analysis of dimeric and trimeric Fe-triazole
compounds~\cite{csd:2005}.  
A hypothetical, but sensible crystal structure of compound {\bf 2} was
built up with the minimum number of atoms per unit cell (72
atoms). The iron-triazole chain itself has $6_3/m\,2/m\,2/c$ symmetry.
A crystal structure with hexagonal symmetry is in principle possible,
{\it e.g.} in space group $P\,6_3/m$, but we chose a structure with
$P\,2_1/m$ symmetry with two formula units per unit cell. A similar
arrangement of chains is also found in a corresponding Cu polymer,
[Cu(hyetrz)$_3$](CF$_3$SO$_3$)$_2 \cdot$H$_2$O~\cite{Garcia03}. (This structure is
triclinic, with space group $P\,\bar{1}$, $Z = 2$, but the deviations
from monoclinic symmetry are probably caused only by the
symmetry-breaking hydroxyethyl ligands and CF$_3$SO$_3^-$ counterions.
Otherwise the symmetry would be $P\,2_1/m$.) In our structure, the
Fe$^{2+}$ ions are located on crystallographic inversion centers,
whereas all triazole units contain a crystallographic mirror plane
between the neighboring nitrogen atoms.  All Fe$^{2+}$ ions are
crystallographically equivalent and we enforce, for simplicity, a
perfectly octahedral environment of the Fe$^{2+}$ ions.


In the design and analysis of the model structures we employ four
distinct classical and {\it ab initio} methods.  For
preoptimization of the model structures we consider a classical  modified
Dreiding force field~\cite{Mayo90} with atomic charges calculated
by the Gasteiger method~\cite{Gasteiger80} as implemented in 
 the Cerius$^2$ package~\cite{cerius:2003}.
We
then perform quantum mechanical first principles calculations within
DFT with three different basis sets, each
for a distinct purpose: The  plane wave {\it ab initio} molecular
dynamics (AIMD) method~\cite{Bloechl94} is used for determination of precise
equilibrium structures. The linearized augmented plane wave (LAPW)
method~\cite{WIEN2K} is used to determine accurate electronic and magnetic
properties and the linearized / Nth order muffin tin orbital
(LMTO/NMTO) methods~\cite{Andersen00} are used to calculate effective Fe Wannier
functions and to understand the low energy excitations of the
 system.

For the AIMD calculation~\cite{Bloechl94} we considered 
a plane wave cutoff of
30~Ryd for the plane wave part of the wave function and  we used the
following sets of (s,p,d) projector functions per angular momentum:
Fe(2,2,2), F(2,2,1), N(2,2,1), C(2,2,1) and H(2,0,0). We employed  a
$(4\times 4\times 4)$ $k$ mesh and the $P\,21/m$ symmetry was preserved during
the relaxation with the help of 131 constraints.

For the LAPW calculations~\cite{WIEN2K},  we employed a
$(6\times 6\times 9)$ $k$ mesh and a value of $R_\mathrm{MT}K_\mathrm{max}=4.2$
that is sufficiently large due to the small radii $R_\mathrm{MT}$ of the
hydrogen atoms.

The NMTO-downfolding~\cite{Andersen00} calculations, which rely on the
self-consistent potentials derived out of LMTO~\cite{Andersen84}
calculations, were carried out with 40 different empty spheres in
addition to atomic spheres to space fill.  The convergence of LMTO
calculations in each case was cross-checked with full potential LAPW
calculations.

All DFT calculations were performed within the
Generalized Gradient Approximation~\cite{Perdew96}.

\begin{figure}
\begin{center}
\includegraphics[width=0.92\textwidth]{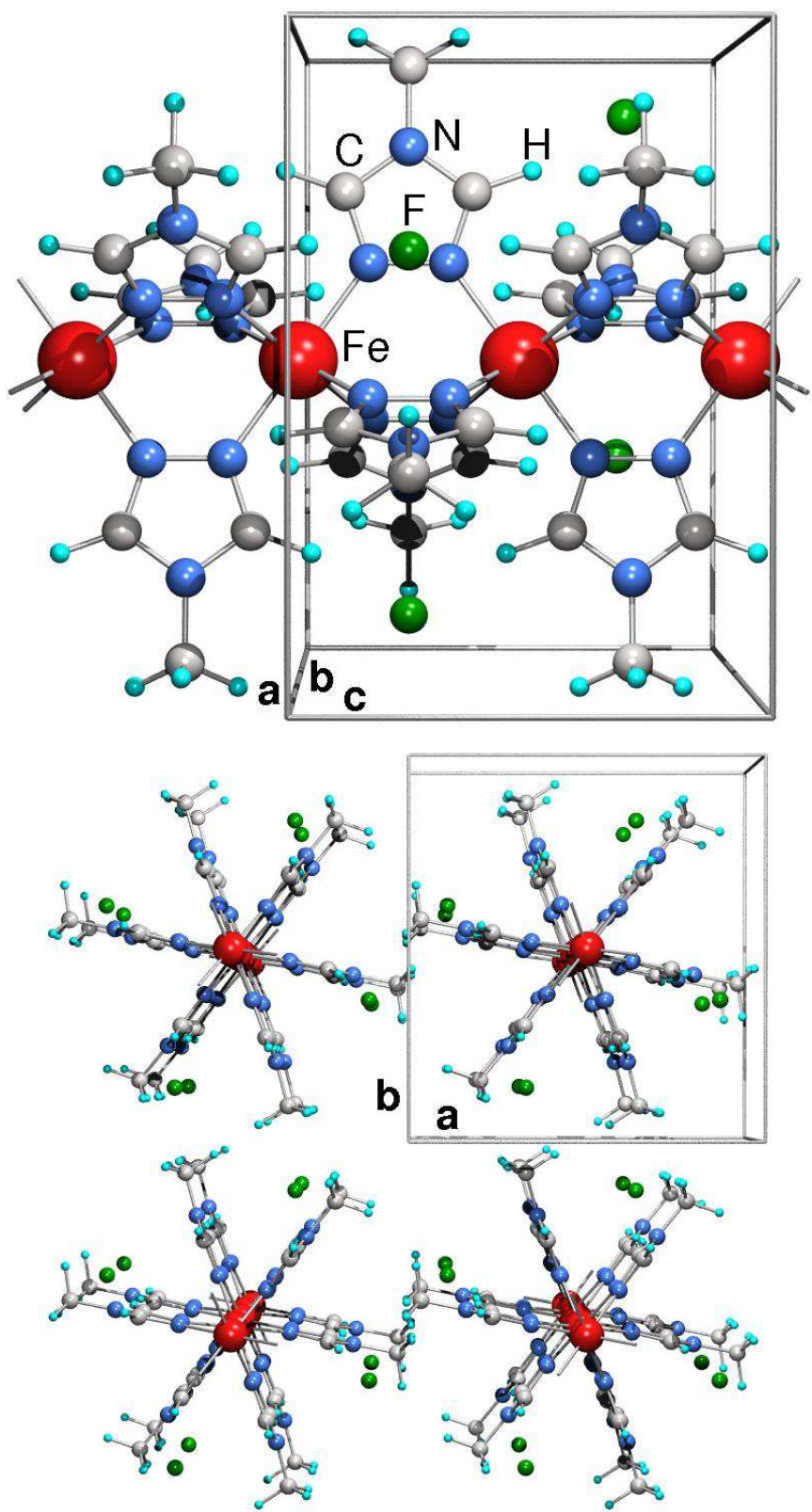}
\end{center}
\caption{
{\bf Simplified Fe(II) triazole structure}. Top: Individual chain of
octahedrally coordinated Fe$^{2+}$ ions. Note the strong linkage
between Fe(II) centers via triple N-N bridges. The orientation of
the FeN$_6$ octahedra is alternating along the chain. Bottom:
Arrangement of polymeric Fe(II) chains in the crystal. }
\label{fig:structure}
\end{figure}

We first proceed with the relaxation
of the model structure with the classical force field within the
$P\,1\,1\,2_1/m$ space group (non-standard setting of $P\,2_1/m$),
keeping fixed the Fe-N distances of the FeN$_6$ octahedra to the
values $d_\mathrm{Fe-N}\in\{2.00$~{\AA}, $2.05$~{\AA}, $2.08$~{\AA},
$2.09$~{\AA}, $2.10$~{\AA}, $2.12$~{\AA}, $2.15$~{\AA},
$2.20$~{\AA}$\}$.  This constraint determines the value of crystal
field splitting from the onset and constitutes our control parameter
as shown below.  As the N-N bond of the triazole (see
Fig.~\ref{fig:triazole}) has experimentally a well defined bond length
of $d_\mathrm{N-N}=1.38$~{\AA}, the choice of a Fe-N distance leads
automatically to lattice parameters linear in $d_\mathrm{Fe-N}$.  The
relative change of the lattice parameter $c$ (the chain direction) is
larger than that for $a$ and $b$.  This means that the volume of the
unit cell also increases linearly from $V=6508$~{\AA}$^3$ at
$d_\mathrm{Fe-N}=2.00$~{\AA} to $V=7317$~{\AA}$^3$ at
$d_\mathrm{Fe-N}=2.20$~{\AA}. It should be noted that the force field
we use is not suitable for relaxing the F$^-$ anion into an
equilibrium position due to the lack of appropriate parameters for
F$^-$ anions.  Therefore, the F$^-$ counter-ion is put into a likely
position and the optimization is left to the next step where we
perform a precise relaxation of the structure with the help of {\it ab
initio} molecular dynamics. This AIMD step is essential as we find
that the force field relaxed structures still show no or very bad
convergence in LAPW, indicating inappropriate positions of at least
some atoms. We relax all unconstrained atoms within AIMD until the
forces that initially are of the order of $1000$~mRyd/Bohr until they
are far below $1$~mRyd/Bohr. More significantly, we converge bond
lengths to within $10^{-3}$~{\AA} and bond angles to within
0.1$^\circ$. We cross check the final AIMD relaxed structures by
calculating the LAPW forces and making sure that they are of the order
of $10$~mRyd/Bohr or less (with the only exception of the difficult to
place counter ions F$^-$ for which we allow $30$~mRyd/Bohr as they do
not play a crucial role for the Fe(II) center interactions). The
stipulation that forces are very low in two fundamentally different
DFT codes is a high standard and gives us confidence that our
conclusions about the electronic and magnetic structure are drawn on a
solid basis. The precision to which we converge bonds and angles means
that we can then proceed to predict interactions between Fe(II)
centers with the necessary precision of a few Kelvin. In
Fig.~\ref{fig:structure} a representative of the resulting structures
is shown. The top panel pictures the chain of FeN$_6$ octahedra with
alternating orientations, and in the bottom panel we demonstrate the
arrangement of the Fe(II) chains in the crystal.

\section{Energy scales}\label{sec:cf}

At the atomic level, two sets of energy scales are responsible
for the LS and HS state of the Fe(II) centers, namely  the crystal-field
splitting and the Hund's exchange and Coulomb interactions.

\begin{figure}
\begin{center}
\includegraphics[width=0.95\textwidth]{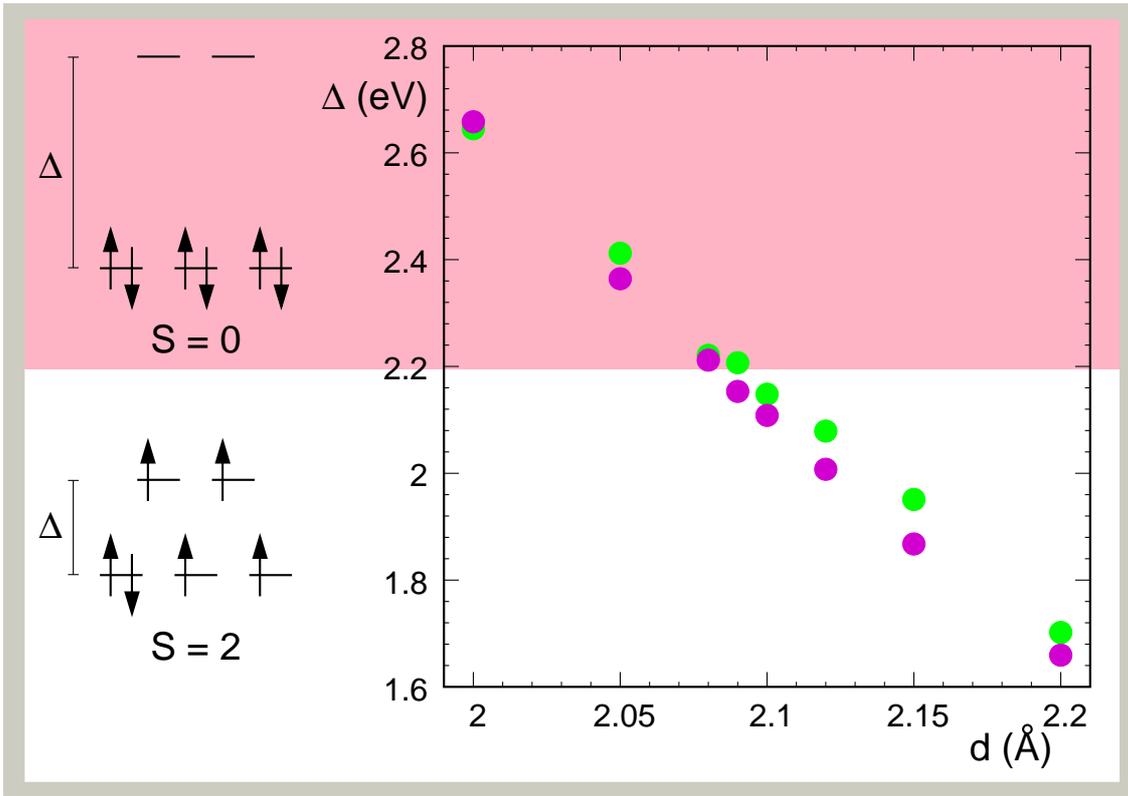}
\end{center}
\caption{
{\bf Crystal field splitting $\Delta$ as function of Fe-N
distances}. The results obtained with LAPW (magenta) and LMTO (green)
agree very well. The pink (white) background symbolizes the respective
colors of the LS (HS) compounds and indicates the $\Delta$ value region
that leads to LS (HS). The atomic Fe $3d$ state diagrams visualize the
relationship between crystal field splitting and spin state. }
\label{fig:crystalfield}
\end{figure}

  The crystal-field splitting, as mentioned above, is being fixed
beforehand in the construction of the structures with given
$d_\mathrm{Fe-N}$ distances. In Fig.~\ref{fig:crystalfield} we show the
dependence of the crystal field splitting on the size of the FeN$_6$
octahedra calculated with LAPW (magenta symbols) and  NMTO (green
symbols). Note the
good agreement between the two calculations.
The crystal field
splitting values are obtained by determining the first moment of both
$t_{2g}$ and $e_g$ densities of states from non-spin polarized
LAPW calculations   and through construction of an
Fe d only Hamiltonian in case of NMTO calculations. 
The atomic state diagrams for Fe 3d schematically
demonstrate the relationship between crystal field splitting $\Delta$
and the spin state ($S=0$ or $S=2$).  The Hund's exchange 
 is taken into account to some extent within the
spin polarized-GGA approach. 

In the extended system, two type of interactions contribute to the phase
transition;
 the phononic excitations 
and
 the exchange interaction $J$, due to
nearest neighbor superexchange between Fe(II) centers which is  typically antiferromagnetic.

The competition between all these energy scales determines the nature of
the phase transition and its cooperativity, as we will discuss in section~\ref{sec:conclusions}. 
The role played by the phonons in driving
the LS-HS transition in the spin-crossover systems has been discussed
at length in terms of elastic
models~\cite{Willenbacher88,Spiering89,Nishino07,Wajnflasz71,Boukheddaden07a}. In
our model calculations, the phononic degrees of freedom are frozen
and we investigate the role of the electronic and magnetic degrees of freedom.

\section{Electronic structure}\label{sec:electronic}

\begin{figure}
\begin{center}
\includegraphics[width=0.95\textwidth]{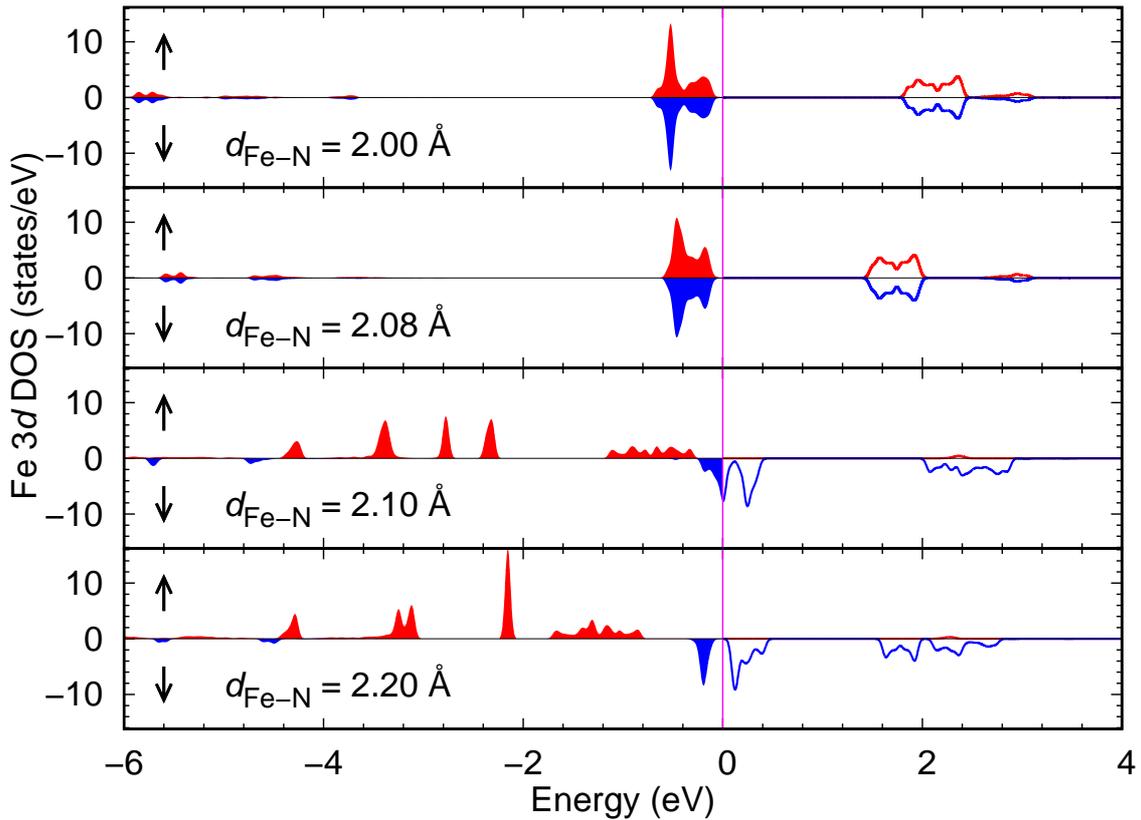}
\end{center}
\caption{
{\bf Spin resolved density of states for selected Fe(II) triazole
structures}. Red and blue colors denote spin up and spin down
species respectively.
The two upper panels for low Fe-N distances show
symmetric density of states corresponding to a low spin ($S=0$)
state. The two lower panels for high Fe-N distances with strong spin
up spin down asymmetry correspond to a high spin ($S=2$) state. }
\label{fig:DOS}
\end{figure}

We performed spin resolved electronic structure calculations for these
systems within DFT by considering the 
GGA~\cite{Perdew96} as exchange-correlation functional
and the LAPW basis. In Fig.~\ref{fig:DOS} we present for four of the
designed structures the projection of the total density of states
(DOS) on the Fe $3d$ orbitals,  which are responsible for the magnetism
of the material. The first and second panels correspond to the
structures with $d_\mathrm{Fe-N}=2.00$~{\AA} and
$d_\mathrm{Fe-N}=2.08$~{\AA} respectively and show a perfect symmetry
between spin up (red) and spin down (blue) density of states and
therefore define a LS state ($S=0$). The occupied states can be
identified as the six $t_{2g}$ states, while the empty states are the four
$e_g$ states. The splitting $\Delta$ between $t_{2g}$ and $e_g$ states
diminishes with increasing $d_\mathrm{Fe-N}$ from $\Delta=2.66$~eV to
$\Delta=2.22$~eV,  respectively (see also
Fig.~\ref{fig:crystalfield}). The DOS behavior completely changes as
the Fe-N distance increases to $d_\mathrm{Fe-N}=2.10$~{\AA} (see
the third panel of Fig.~\ref{fig:DOS}).
Now spin up (red) $t_{2g}$ and $e_g$ states  are completely filled, and
spin down (blue) $t_{2g}$ states  show only a partial filling with one
electron. The imbalance between up and down electron numbers is
$n_{\uparrow}-n_{\downarrow}=4$ which corresponds to the HS state
($S=2$). This situation remains if  the Fe-N
distance increases further  to $d_\mathrm{Fe-N}=2.20$~{\AA}, only the splitting $\Delta$
between $t_{2g}$ and $e_g$ diminishes.

Thus, by carefully preparing a series of model structures that
correspond to the LS and HS sides of the spin crossover transition we
manage to microscopically describe the LS-HS transition which occurs
between the structures with $d_\mathrm{Fe-N}=2.08$~{\AA} ($S=0$) and
$d_\mathrm{Fe-N}=2.10$~{\AA} ($S=2$).

\begin{figure}
\begin{center}
\includegraphics[angle=-90,width=0.8\textwidth]{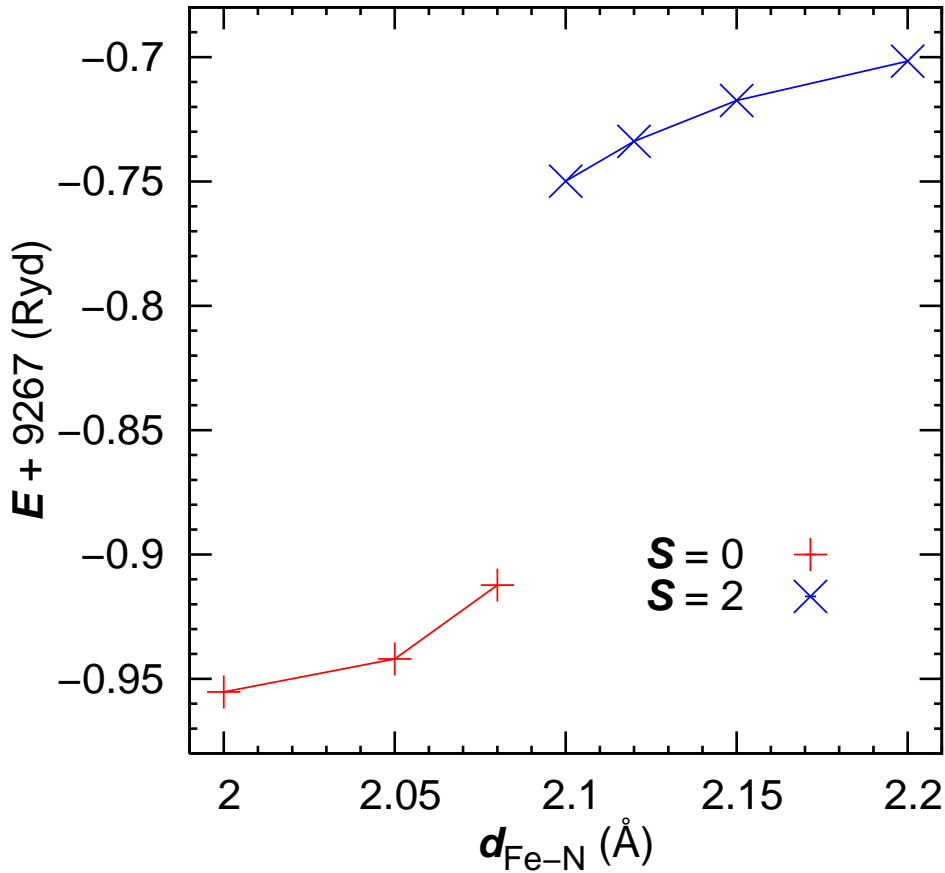}
\end{center}
\caption{
{\bf  Ground state energies for the set of
 model structures obtained within  sp-GGA with
the LAPW basis }. }
\label{fig:energies}
\end{figure}

In order to quantify energetically the HS-LS spin transition, we show
in Fig.~\ref{fig:energies} the total LAPW electronic energies obtained
within the spin-polarized GGA (sp-GGA) approach. We note that there is
a discontinuous jump between the LS (S=0) energies and the HS (S=2)
energies. The relative electronic energy differences between the HS
and LS systems $\frac{E_{el}^{(HS)}-E_{el}^{(LS)}}{E_{el}^{(HS)}}$ is
about $10^{-5}$ which agrees with the relative energy estimates for
spin crossover molecular systems~\cite{Paulsen04}. Since the LS-HS
phase transition occurs between structures $d_\mathrm{Fe-N}=2.08$~{\AA}
and $d_\mathrm{Fe-N}=2.10$~{\AA}, we designed one more structure with
$d_\mathrm{Fe-N}=2.09$~{\AA} in order to probe the sharpness of the
transition. While there are indications that this structure might
represent an intermediate magnetic state with $S \approx 1.5$ per
Fe(II) center, we do not include it in Fig.~\ref{fig:energies} as it
is very hard to converge.  This result indicates that the spin
crossover transition in polymer systems occurs in a very narrow range
of crystal field splittings, i.e. is very sharp, as observed
experimentally~\cite{Kahn98}.

\section{Exchange interaction and magnetic properties}\label{sec:exchange}

\begin{figure}
\begin{center}
\includegraphics[angle=-90,width=0.9\textwidth]{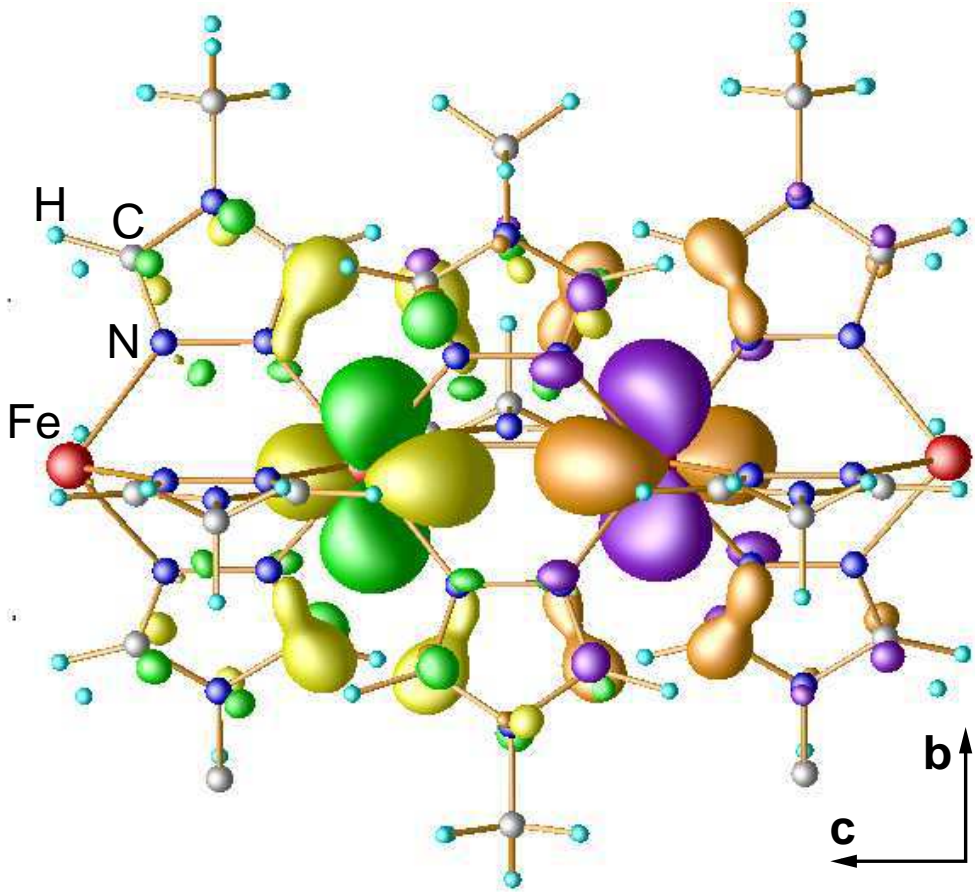}
\end{center}
\caption{
{\bf Wannier functions for the Fe-Fe interactions}. The chosen example
shows two Wannier functions with $3d_{xy}$ symmetry on two
neighboring Fe(II) centers. The effective interaction (superexchange
coupling) between Fe(II) centers depends on the degree and relative
sign of the overlap of the Wannier function tails on the pyrazole
ring.  }
\label{fig:wannier}
\end{figure}

For the magnetic behavior of this system, we derive first from the DFT
electronic calculations a Hamiltonian which describes the effective
interaction between Fe(II) centers. The NMTO-based downfolding 
method, 
designed to pick up selectively the low-energy bands from the whole
group of LDA/GGA bands of a compound,   has been used to construct the
Fe $3d$ only Hamiltonian of the Fe-triazole compounds. This method has proven to provide reliable information on the low energy properties of inorganic~\cite{Valenti01,SahaDasgupta05,Rahaman07} and organic~\cite{Salguero07} transition metal compounds.
The tight-binding basis in which these Hamiltonians are constructed form
the set of "effective" functions, which span the Hilbert space of the
Wannier functions corresponding to the low-energy
bands. Fig.~\ref{fig:wannier} shows the plot of one member of such a
set, namely the downfolded Wannier function corresponding to
Fe~$3d_{xy}$. In the figure, two such Fe~$3d_{xy}$ Wannier functions
have been placed at two neighboring Fe sites. While the central part
of such an effective function has the Fe $3d_{xy}$ symmetry, the tails
of the function are shaped according to integrated out degrees of
freedom in the system, like C~$sp$, N~$sp$, F~$sp$ and H~$s$. As is
evident from the plot, substantial weight of these tails resides on
neighboring triazole rings. The presence of these tails hints to an
enhanced electronic hybridization between the adjacent Fe$^{2+}$ ions,
contributing to the cooperative nature of the HS-LS transition. Out of
these calculations we can estimate the various hopping matrix
elements, $t$, between the $d$ orbitals of adjacent Fe(II) centers.
The values of these hopping parameters range between 1 meV to 80 meV
quantifying the strength of the various interaction paths between neighboring
$d$ Fe(II) orbitals.

In order to get estimates of the magnetic superexchange coupling
constants $J$  between neighboring Fe(II) centers
for the HS Fe-triazole structures, we considered two
approaches: \\
i) Total energies.  We calculated within sp-GGA total energies of
ferromagnetic and antiferromagnetic Fe$^{2+}$ spin
configurations. Considering a spin-Hamiltonian $H= J S_i S_{i+1}$
between nearest neighbors Fe$^{2+}$ spins $S_i$, the ferromagnetic and
antiferromagnetic energies for two Fe$^{2+}$ ions in the unit cell of
the Fe-triazole are given by $E_\mathrm{FM} = 8J$ and
$E_\mathrm{AFM}=-8J$.  Within sp-GGA $E_\mathrm{FM} - E_\mathrm{AFM} =
33$~meV for the $d_\mathrm{Fe-N}=2.20$~{\AA} structure and therefore $J
\approx 2.1~meV =24 ~K$.
ii) Perturbation theory~\cite{Anderson78} where $J$ can be
obtained in terms of the hopping parameters $t$ between Fe(II) centers
and the onsite Coulomb repulsion $U$ as $ J \approx 4 t^{2}/U$. For the HS $d_\mathrm{Fe-N}=2.20$~{\AA} structure, the significant $t$
obtained within the downfolding method is 48 meV, and for $U$ = 4-5 eV
this gives $J \approx 22 K$
which is very similar to the value obtained
with the difference of total energies.

\begin{figure}
\begin{center}
\includegraphics[width=0.95\textwidth]{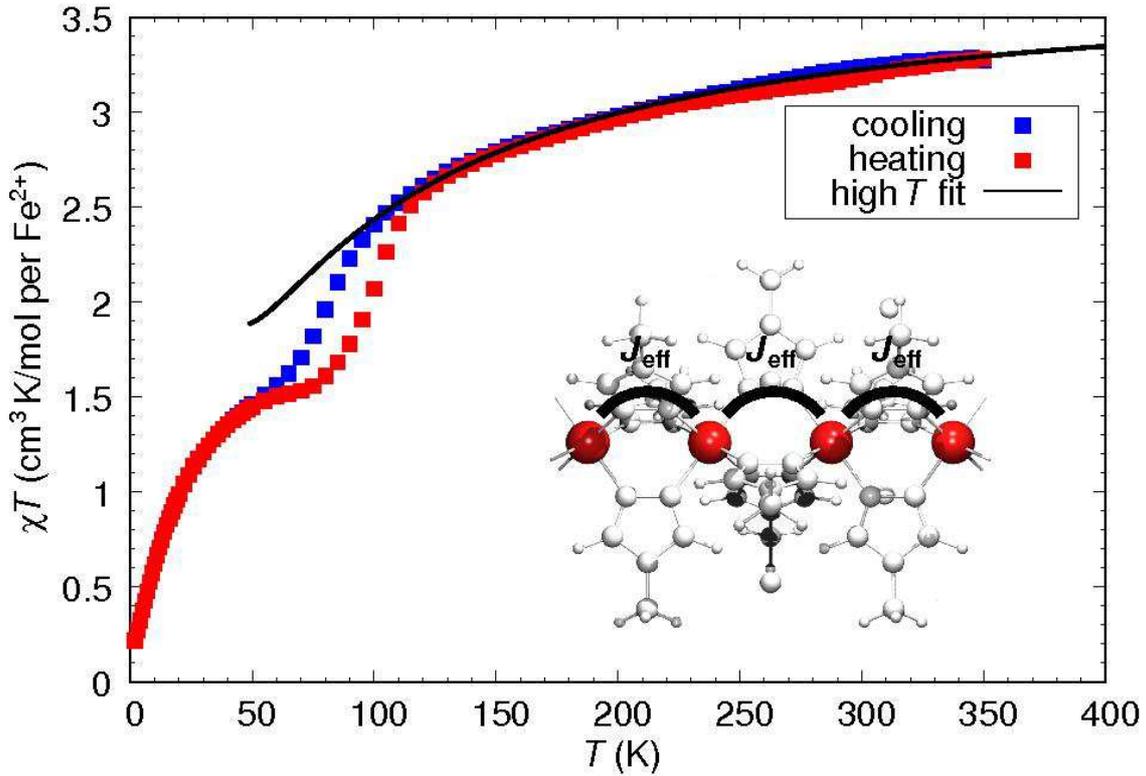}
\end{center}
\caption{
{\bf Measured susceptibility.} The solid line corresponds to a high-T
fit as explained in the text. The susceptibility measured while
lowering (increasing) the temperature shows a hysteresis of 20~K at
the transition temperature of 80~K.  }
\label{fig:susc}
\end{figure}

The results of our model calculations can be now compared with the
magnetic properties obtained on the real samples of polymeric
Fe[(hyetrz)$_3$](4-chlorophenylsulfonate)$_2 \cdot
3$H$_2$O. Variable-temperature magnetic susceptibility measurements in
the temperature range $2-350$~K and magnetic fields $0.02-0.2$~T were
carried out on powder samples of Fe(II) triazole using a Quantum
Design SQUID magnetometer MPMS-XL.  In Fig.~\ref{fig:susc} we show the
susceptibility $\chi$ measurements where $T\chi$ has been plotted
versus $T$. Our sample shows hysteresis at $T= 80$~K with a width of
20~K. Since this system consists of spin $S=2$ Fe(II) chains with weak
interchain interactions, we have analyzed the magnetic susceptibility
in the frame of a spin $S=2$ Heisenberg chain model.

 Note that the Fe(II) triazole SCP systems 
are Haldane $S=2$ chains.  A Haldane gap is expected to exist between
the ground state and first excited states.  The reason for not
observing this gap, is that the transition temperature at which the
HS-LS transition happens is higher in energy than the Haldane gap
energy and therefore, the system goes into the non-magnetic $S=0$
phase before the gap in the $S=2$ chain can be observed.

The susceptibility  of an $L$-site  chain  is 
given by: 
\begin{equation}
\chi_L = \frac{g^2\mu_B^2}{k_\mathrm{B} T} 
\frac{Tr\left[(\sum_{i=1}^L S_i^Z)^2 e^{-\beta H}\right]}{Tr\left[
e^{-\beta H}\right]}
\label{eq:susceptibility}
\end{equation}
where  $H$ is the Heisenberg
Hamiltonian $H=J S_i S_{i+1}$, 
$\mu_B$ is the Bohr magneton, $g$ is the gyromagnetic
factor, $k_\mathrm{B}$ is the Boltzmann constant, $T$ is temperature and
$S^z_i$ is the $z$-component of the spin on site $i$.
In the thermodynamic limit,
the bulk susceptibility at high temperatures can
 be obtained as a series expansion in $1/T$:
\begin{eqnarray}
\frac{\chi J}{g^2\mu_B^2}= \frac{1}{3}\left[\frac{S(S+1)J}{k_\mathrm{B} T}\right
] - 8 \Bigl(\frac{J}{k_\mathrm{B} T}\Bigr)^2 + 16
\Bigl(\frac{J}{k_\mathrm{B} T}\Bigr)^3 + O(\Bigl( \frac{J}{k_\mathrm{B}
T}\Bigr)^4) \label{eq:highTsusc}
\end{eqnarray} 
For $k_\mathrm{B} T/J > S(S+1)$  Eq.~\ref{eq:highTsusc}
compares very well to QMC data for
spin $S=2$ chains~\cite{Yamamoto96}.

The fit of Eq.~(\ref{eq:highTsusc}) to the measured susceptibility of
Fig.~\ref{fig:susc} is best for $g=2.2$ and $J=11$~K.

The $J$ values obtained from our {\it ab initio} calculations ($J$
$\approx 24 K$) are slightly larger than the $J$ extracted from the
susceptibility data ($J = 11$~K) on the real sample, but remain in the
same order of magnitude. Considering that i) we performed the
calculations in a model structure and ii) the experimental
measurements are affected by the quality of the samples, we can
conclude that the comparison is quite good and the designed structures
are reliable.

\section{Discussion}\label{sec:conclusions}

One important topic of this work is the analysis of the various energy
scales that contribute to the cooperativity of the HS-LS transition in
SCP systems.  In our calculations we froze the elastic degrees of
freedom and concentrated on the electronic and magnetic properties for
which we have quantified the corresponding parameters.  A comparison
with the elastic coupling constants estimated from Ising-like
models~\cite{Boukheddaden07a} (in such an approach, a material with a
transition temperature of $T=80$~K would be described by elastic
interactions of $J_{elastic} \approx 20-30 $ K) shows that in
one-dimensional Fe(II) triazole, the elastic coupling constants
$J_{elastic}$ are of equal importance as the magnetic exchange for
explaining the spin crossover transition, and the cooperativity should
be understood as an interplay between elastic properties and magnetic
exchange. When we cool the system from the HS state toward the HS-LS
transition temperature, the elastic coupling tends to drive the system
to the LS state, while the magnetic exchange tends to keep up the
magnetic state for a larger temperature range (see
Fig.~\ref{fig:susc}). In comparison, in the heating process, the
magnetic exchange is initially absent (LS) and therefore the elastic
interaction (vibrational phonons) initially drives the transition which has its on-set at a
higher temperature than in the cooling process. The width of the
transition between cooling and heating (hysteresis) is therefore
enhanced by the magnetic interaction.

A fundamental difference between the polymeric systems we are dealing
with in this work, and molecular bi- (tri-, tetra-, \dots) nuclear Fe
systems is the connectivity between the Fe(II) centers.  While the
molecular systems~\cite{Postnikov06} form isolated clusters of Fe(II) centers and
therefore there is no strong connectivity between clusters, the
polymers have important nearest neighbor interactions in the
thermodynamic limit. This implies that for the polymers, the magnetic
superexchange is not restricted to the cluster as in the molecular
systems, but rather becomes important for the nature of the HS-LS
phase transition.  In Ref.~\cite{Desroches06} various estimates
of the magnetic J for molecular systems have been given.  The values
range between 4-6 K.  The values we estimated for the present polymers
are larger, between 11-24~K and in the energy range of the elastic
constants, which indicates that the cooperativity in these SCP systems
is most likely significantly enhanced by the exchange interactions.

In conclusion, this work presents an efficient route to prepare
reliable model structures for microscopic investigations and provides
a new interpretation about the origin of the parameters underlying
traditional theoretical approaches for the polymeric spin-crossover
materials. The next step of this study will be the inclusion of
lattice dynamics (phonons) in the {\it ab initio} calculations which
is planned in our future work.

\ack

The authors thank Chunhua Hu, J\"urgen Br\"uning and Jens K\"uhne for
the syntheses and P. G\"utlich for stimulating discussions. We would like to acknowledge the German Science
Foundation (DFG) for financial support through the FOR~412, TRR/SFB~49 and
Emmy Noether programs. T.S.D. thanks the MPG-India partnergroup
program for the collaboration. We gratefully acknowledge support by
the Frankfurt Center for Scientific Computing.

\end{document}